\documentclass[pra,twocolumn,superscriptaddress,showpacs]{revtex4}
\usepackage[english]{babel}
\usepackage[dvips]{graphicx}
\usepackage{amssymb}
\usepackage{amsmath}
\usepackage{color}
\begin{document}

\title{Single-qubit lasing in the strong-coupling regime}

\author{Stephan Andr\'e}
\affiliation {Institut f\"ur Theoretische Festk\"orperphysik,
      Karlsruhe Institute of Technology, 76128 Karlsruhe, Germany}
\affiliation {DFG Center for Functional Nanostructures (CFN),
      Karlsruhe Institute of Technology, 76128 Karlsruhe, Germany}
\author{Pei-Qing Jin}
\affiliation {Institut f\"ur Theoretische Festk\"orperphysik,
      Karlsruhe Institute of Technology, 76128 Karlsruhe, Germany}
\author{Valentina Brosco}
\affiliation {Institut f\"ur Theoretische Festk\"orperphysik,
      Karlsruhe Institute of Technology, 76128 Karlsruhe, Germany}
\affiliation{Dipartimento di Fisica, Universit\`a ``La Sapienza'', P.le
A. Moro 2, 00185 Roma, Italy}
\author{Jared H.~Cole}
\affiliation {Institut f\"ur Theoretische Festk\"orperphysik,
      Karlsruhe Institute of Technology, 76128 Karlsruhe, Germany}
\author{Alessandro Romito}
\affiliation {Institut f\"ur Theoretische Festk\"orperphysik,
      Karlsruhe Institute of Technology, 76128 Karlsruhe, Germany}
\affiliation {Freie Universit\"at Berlin, Fachbereich Physik,
      Arnimallee 14, 14195 Berlin, Germany}
\author{Alexander Shnirman}
\affiliation{Institut f\"{u}r Theorie der Kondensierten Materie,
Karlsruhe Institute of Technology, 76128 Karlsruhe, Germany}
\affiliation {DFG Center for Functional Nanostructures (CFN),
      Karlsruhe Institute of Technology, 76128 Karlsruhe, Germany}
\author{Gerd Sch\"on}
\affiliation {Institut f\"ur Theoretische Festk\"orperphysik,
      Karlsruhe Institute of Technology, 76128 Karlsruhe, Germany}
\affiliation {DFG Center for Functional Nanostructures (CFN),
      Karlsruhe Institute of Technology, 76128 Karlsruhe, Germany}

\date{\today}

\begin{abstract}
Motivated by recent ``circuit QED'' experiments we study the lasing transition 
and spectral properties of single-qubit lasers. 
In the strong coupling, low-temperature regime quantum fluctuations dominate 
over thermal noise and strongly influence the linewidth of the laser. 
When the qubit and the resonator are detuned, amplitude and phase fluctuations 
of the radiation field are coupled, and the phase diffusion model, commonly used to
describe conventional lasers, fails. We predict pronounced effects near the lasing transition, 
with an enhanced linewidth and non-exponential decay of the correlation functions.
% These features show up already, although less pronounced, at intermediate coupling strength.
We cover a wide range of parameters by using two complementary approaches,
one based on the Liouville equation in a Fock state basis, covering 
arbitrarily strong coupling but limited to low photon numbers,
the other based on the coherent-state representation, covering large photon numbers
but restricted to weak or intermediate coupling.
\end{abstract}

\maketitle

\section{Introduction}

A  strong,  coherent
coupling between a superconducting qubit and an electrical resonator
was first realized experimentally by Wallraff \emph{et al.} and
Chiorescu \emph{et al.}  \cite{wallraff04,chiorescu}.  Their work stimulated
substantial  theoretical
\cite{blais04,rodrigues07-1,hauss08,marthaler08,zhirov,gambetta,li} and
experimental activities
\cite{naik,johansson,schuster,sillanpaa,astafiev,grajcar,hofheinz}
devoted to the study of further quantum electrodynamic effects in electric circuits.
In these ``circuit QED'' setups a superconducting qubit plays the
role of an artificial atom, while the radiation field is replaced by the
modes of an electric resonator.
In some of these experiments a strong enhancement
of the resonator field and lasing were observed  \cite{astafiev,grajcar}.

In contrast to conventional lasers where many atoms are 
coupled weakly to the light field in an optical interferometer, in the
single-qubit laser a single superconducting qubit is coupled strongly to the microwave field of the resonator circuit.
Furthermore, typical circuit QED setups operate at low temperatures, where
thermal noise is weak and quantum fluctuations, arising from the qubit-resonator coupling,
become dominant \cite{Lugiato,Drummond,Wallraff2010}.

In this work we study the lasing transition and the spectral properties of single-qubit lasers
focusing on the regime of strong qubit-resonator
coupling at low temperatures.
In this regime we find qualitatively new behavior of the laser linewidth.
In earlier work~\cite{brosco1,brosco2} we observed that the linewidth
depends in a non-monotonous way on the coupling strength:
approaching the lasing transition from weak coupling one observes the well-known linewidth narrowing.
However, for stronger coupling the linewidth of the single-qubit laser increases again,
and the lasing state deteriorates, a behavior not observed in conventional lasers.

If the qubit and the laser are detuned from resonance we find an even more surprising behavior:
The linewidth, which generally increases with detuning, is strongly enhanced 
near the lasing transition, and for strong coupling or low damping rate of the resonator shows a local maximum 
as a function of detuning.
In this regime, the correlation functions of the resonator decay non-exponentially in time.
As we demonstrate below, this behavior is due to 
the coupling between phase and amplitude fluctuations,
which are neglected in the commonly used phase diffusion model \cite{Scully}.

In order to cover a wide range of parameters we analyze the properties of the laser using two approaches.
One is based on a numerical analysis of the Liouville equation in a basis of Fock states.
It allows for arbitrarily strong coupling but is limited by the size of the basis which we can handle.
This puts constraints on the thermal photon number and the quality factor of the resonator.
The other is based on the coherent-state representation.
It covers large photon numbers (and thus low damping rate of the resonator) 
but fails in the strong coupling regime.
The results of the two approaches coincide well in an interesting overlapping parameter regime.

The paper is organized as follows:
In Sec.\ \ref{sec:model}, we describe the model and in Sec.\ \ref{sec:methods}
the two approaches employed to solve the master equation.
Stationary properties, such as the average photon number are evaluated in Sec.\ \ref{sec:transition}.
In the central part of this paper, Sec.\ \ref{sec:linewidth},
we investigate the spectral function and the linewidth near the lasing transition
in the frame of the two approaches.
The consequences of  amplitude fluctuations and the failure of the phase diffusion model
are discussed.
We draw conclusions in Sec.\ \ref{sec:discussion}.

\section{The model}\label{sec:model}

The single-qubit laser realized by Astafiev et al.~\cite{astafiev} consists of 
a Cooper pair box (CPB)~\cite{CPB}, characterized by the charging energy scale 
$E_{\rm ch}$ and the Josephson coupling energy $E_{\rm J}$, which is 
coupled capacitively to a superconducting coplanar waveguide resonator
with frequency  $\omega_{\rm R}$.
In the regime typically explored in the experiments, only two charge states of the CPB, 
indicated as $|0\rangle$ and $|2\rangle$ and differing by one Cooper pair, 
are coherently  and (near) resonantly coupled to the cavity. 
In this situation, the system can be modelled as a two-level system (qubit) 
coupled to a harmonic oscillator with Hamiltonian \cite{blais04}
\begin{equation}
H = -\frac{1}{2}(E_{\rm ch} \hat\tau_z
 +E_{\rm J}\hat\tau^{}_x)
 +\hbar\omega_{\rm R} a^\dag a^{}
 +\hbar g_0 \hat\tau_z (a+a^\dag) \, .
\label{eq:H-full}
\end{equation}
Here we set $\hat\tau_z=|0\rangle\langle0|-|2\rangle\langle2|$, 
$\hat\tau_x=|0\rangle\langle2|+|2\rangle\langle0|$, while $a$ and $a^{\dag}$ are the annihilation and
creation operators of photons in the resonator.
The qubit-resonator coupling strength is denoted by  $g_0$.
In the considered single-qubit laser experiments, the qubit level spacing  and $\omega_{\rm R}$
are in the range of 10~GHz (i.e., microwave frequencies)
while the strength of the coupling  $g_0$ reaches 100~MHz. 
This is much stronger than usual for conventional lasers but still small enough to allow for the rotating-wave approximation.
Switching to the qubit's eigenbasis we can recast 
the Hamiltonian in the  Jaynes-Cummings form \cite{JC},
\begin{equation}
H_{\rm JC} = \frac{1}{2}\hbar
\omega_{\rm Q} \sigma_z + \hbar \omega_{\rm R} a^{\dagger} a
 +\hbar g \left( \sigma_+ a + \sigma_- a^{\dagger}\right) \, .\label{eq:H-JC}
\end{equation}
Here we introduce new Pauli matrices, $\sigma_{x,y,z}$  and  $\sigma_{\pm}=\sigma_x\pm i\sigma_y$.
Furthermore, the the qubit's level spacing $\hbar\omega_{\rm Q}=\sqrt{E_{\rm ch}^2+E_{\rm J}^2}$ and the effective coupling strength $g = \sin\theta ~ g_0$ with $\theta = -\arctan(E_{\rm J}/ E_{\rm ch})$ 
have been introduced.
In the following we allow the qubit and the resonator to be detuned by $\Delta = \omega_{\rm Q}-\omega_{\rm R}$. 
Although we consider the experiment of Ref.\ \cite{astafiev} as a motivation for this analysis,
our model also applies to many further circuit QED experiments in the lasing regime\ \cite{hauss08}.

As discussed in detail in Refs.~\cite{brosco2,didier}, the pumping mechanism implemented by Astafiev et al.~\cite{astafiev} depends on current injection in a superconducting single-charge transistor and 
actually involves three charge states. Besides the states $|0\rangle$ and $|2\rangle$, also a one-excess-electron charge state is incoherently coupled to the system and excited via an external voltage bias. 
For the present discussion it is sufficient to model this driving simply by 
adding an incoherent excitation term to the qubit's dynamics.
%assuming a negative effective temperature of the qubit's bath.

In this case, within the usual Markov approximation, for weak system-environment coupling and
using the secular approximations, we can analyze the dynamics of a single-qubit laser in the frame of a Bloch-Redfield master equation \cite{fano,ct}.
The reduced qubit-resonator density matrix $\rho$ obeys the master equation in Lindblad form \cite{gardiner}, which in the laboratory frame reads 
\begin{equation} \label{eq:MELiouville}
\dot \rho=-\frac{i}{\hbar}\left[H_{\rm JC}, \rho\right]
+\mathcal L_{\rm Q}\, \rho+\mathcal L_{\rm R} \, \rho \, .
\end{equation}
Here the Liouville super-operators $\mathcal L_{\rm R}$ and $\mathcal L_{\rm Q}$  account for 
the resonator's and the qubit's dissipative processes,
\begin{eqnarray}\label{eq:MEdampOsc}
\mathcal L_{\rm R} \, \rho=& &\frac{\kappa}{2} (N_{\rm th}+1)
\left(2a \rho a^{\dagger} -a^{\dagger} a\rho -\rho
a^{\dagger} a \right) \nonumber \\
& & + \frac{\kappa}{2} N_{\rm th} \left( 2 a^{\dagger}
\rho a -a a^{\dagger}\rho -\rho aa^{\dagger} \right),
\end{eqnarray}
and %
\begin{eqnarray}\label{eq:MEdampQ}
\mathcal L_{\rm Q} \, \rho= & & \frac{\Gamma_{\varphi}^*}{2} \left(\sigma_z\rho\sigma_z - \rho\right) \nonumber \\
& & + \frac{\Gamma_{\downarrow}}{2} \left(2 \sigma_- \rho \sigma_+ -\rho\sigma_+\sigma_- - \sigma_+ \sigma_-\rho \right) \nonumber \\
& & + \frac{\Gamma_{\uparrow}}{2} \left(2 \sigma_+ \rho\sigma_- -\rho\sigma_-\sigma_+ - \sigma_-\sigma_+\rho \right) \, .
\end{eqnarray}
The dissipative dynamics of the resonator depends
on the bare damping rate $\kappa$ and the thermal photon number $N_{\rm th}$,
while the qubit's dynamics is described by excitation, relaxation,
and pure dephasing with rates $\Gamma_{\uparrow}$, $\Gamma_{\downarrow}$, and $\Gamma_{\varphi}^*$, respectively.
For later use, we also introduce the inverse of the $T_1$ time, $\Gamma_1=\Gamma_{\uparrow}+\Gamma_{\downarrow}$,
the total dephasing rate $\Gamma_{\varphi}=\Gamma_1/2 + \Gamma_{\varphi}^*$, as well as the 
bare population inversion of the qubit $\tau_0=\left(\Gamma_{\uparrow}-\Gamma_{\downarrow}\right)/\Gamma_1$.

As mentioned in the Introduction, in single-qubit lasers the 
lasing transition and state depend on the coupling strength $g$ and 
the qubit-resonator detuning $\Delta$.
In order to characterize the lasing state in these systems we, therefore,
study the stationary properties of the radiation field, such as average  photon number 
and photon number fluctuations as a function of these parameters. 
To fully characterize the lasing state we further investigate 
the emission spectrum of the field,
\begin{eqnarray}
 S_{\rm e}(\omega) = \int^\infty_{-\infty}\!\!\! dt\, e^{-i\omega t}
 \langle a^\dag(t) a(0) \rangle \, ,
\label{eq:Spectrum-Def}
\end{eqnarray}
and the linewidth of the laser radiation. 
This allows us to describe the transition between the incoherent and 
the coherent state of the resonant cavity and to draw a phase diagram of the lasing state.

\section{Methods}\label{sec:methods}

We address the lasing state and the resonator spectrum
by two different methods, by a direct numerical diagonalization
of the master equation for the density matrix and by
a Fokker-Planck equation approach. The two methods
cover different parameter regimes and provide
complementary physical pictures.

\subsection{Direct integration of the Liouville equation (DILE)}
\label{subsec:DirectInt}

After projecting on the Fock states basis, we can recast the master equation (\ref{eq:MELiouville}) in  
vector form
\begin{equation}
 \dot{\vec{\rho}} = G \vec{\rho} \, .
\label{eq:Num-rho}
\end{equation}
The reduced density matrix $\rho$ is arranged as a vector $\vec{\rho}$,
and $G$ is a superoperator acting in the space of the system operators.
This form is convenient for the numerical evaluation of both static and spectral properties of the field.

From the stationary solution of the master equation
$G\vec{\rho}_{\rm s}=0$ 
we obtain equilibrium
values such as, e.g., the photon number in the resonator, $\langle n
\rangle = \textrm{Tr}\lbrace a^{\dagger} a \rho_{\rm s}\rbrace$. 
From Eq.~(\ref{eq:Num-rho}) we
can also derive expressions for
time-dependent correlation functions \cite{Ginzel}, e.g.,
$\langle a^{\dagger}(t)a(0)\rangle$.
In order to do so, we employ the quantum regression theorem \cite{gardiner,Carmichael}
\begin{equation}
 \langle a^{\dagger}(t)a(0)\rangle = \textrm{Tr}\lbrace a^{\dagger} e^{G t} a \rho_{\rm s} \rbrace.
\label{eq:Num-corr}
\end{equation}

To proceed it is convenient to diagonalize the superoperator $G$ and express the product $a\rho_{\rm s}$ in terms of eigenvectors of $G$, $a\rho_{\rm s} = \sum_k c_k \vec{v}_k$. Acting on the equation with the exponential $\exp(Gt)$ leads to 
$$e^{G t} a \rho_{\rm s} = \sum_k c_k e^{\lambda_kt} \vec{v}_k , $$
where $\lambda_k$ is the eigenvalue corresponding to the eigenvector $\vec{v}_k$. Once the eigenvalues $\lambda_k$ and expansion coefficients $c_k$ are known, we can easily obtain the correlation function for all times $t$.

We solved Eqs.~(\ref{eq:Num-rho}) and (\ref{eq:Num-corr}) numerically with results to be presented below.
For this purpose we truncate the Hilbert space of the resonator to a finite
 number $N$ of photon number states.
In this case the superoperator $G$ is  represented by a $4N^2 \times 4N^2$-matrix, growing fast with $N$, which limits the method in our case to $N\le 30$.
This  puts constraints to the values of rates $\kappa$ and $\Gamma_1$. As will be shown in Sec.~\ref{sec:transition}, the average photon number $\langle n \rangle$ in the resonator obeys the relation $\langle n \rangle \lesssim \Gamma_1/(2 \kappa)$. Hence the damping rate $\kappa$ should not be too small,i.e., $\kappa \gtrsim \Gamma_1/N$. 
On the other hand, for the calculation of average steady state values, as well as for the linewidth  exactly at resonance ($\Delta=0$), the calculations reduce to solving a system of linear equations. In these cases, we can cover a much higher number of photon number states, $N\lesssim 200$.

\subsection{Fokker-Planck equation (FPE)}\label{sec:FPE}

Alternatively, following the route outlined in Ref.~\cite{gardiner}, 
we can derive a Fokker-Planck equation for the single qubit-laser
in the coherent state representation. 
In this case we represent the qubit-oscillator density matrix  as 
 $\rho(t)=\int d^2\alpha \, \tilde \rho(\alpha,t)\, |\alpha\rangle\langle\alpha|$, 
 where $\tilde{\rho}(\alpha,t)$ is a $2\times2$ matrix in the qubit basis states.
Substituting this expansion into Eq.~(\ref{eq:MELiouville}) we obtain 
a master equation for the operator $\tilde{\rho}(\alpha,t)$.
Its trace,
$\textrm{Tr}_Q \left[\tilde{\rho}(\alpha,t)\right] \equiv P(\alpha,t)$,
 yields the probability for the radiation field to be in a coherent state $|\alpha\rangle$.

In typical experimental configurations \cite{astafiev},
the dynamics of the qubit are much faster than that of the resonator.
As a result the qubit decays on short time scales 
to a quasi-steady state, which still depends on the slowly varying
state of the radiation field $\alpha(t)$.
We can adiabatically eliminate the qubit's degrees of freedom via a
projective technique \cite{gardiner}.
The resulting FPE contains a
noise term arising from the qubit fluctuations. 
Generally, this noise term contains all
orders of derivatives $\partial/\partial\alpha$ and
$\partial/\partial\alpha^*$. If the coupling  is weak
compared to the dephasing rate of the qubit, $g \ll \Gamma_{\varphi}$,
the noise term can be truncated to second order derivatives,
and the resulting FPE in the rotating frame reads \cite{gardiner}
\begin{eqnarray}\label{eq:FP}
\frac{\partial P}{\partial t}
 &=& \frac{\kappa}{2}\left\{
 \frac{\partial}{\partial\alpha}
 \left[\alpha-\frac{\alpha(1-i\Delta/\Gamma_\varphi) C_0}
 {(1+\Delta^2/\Gamma_\varphi^2)(1+|\alpha|^2/n_0)}
 \right] \right.
            \nonumber \\
 && \left. +\frac{\partial^2}{\partial\alpha\partial\alpha^*}
 ( N_{\rm th} + Q)
 +\frac{\partial^2}{\partial\alpha^2} R
 + {\rm c.c.}
 \right\} P \, .
\end{eqnarray}
Here $C_0 = 2g^2\tau_0/(\kappa \Gamma_{\varphi})$ and
$n_0 = \Gamma_1\Gamma_\varphi/(4g^2)$. The resonator-qubit
coupling leads to the noise terms
\begin{eqnarray}
 Q &=&
\frac{g^2}{\kappa}~\textrm{Re}\Bigl[\int^{\infty}_0
dt(\langle\sigma_+(t)\sigma_-(0)\rangle_{\rm{q}}
 -\langle\sigma_+\rangle_{\rm{q}}\langle\sigma_-\rangle_{\rm{q}} ) \Bigr],
           \nonumber \\
R &=& -\frac{g^2}{\kappa} \int^{\infty}_0 dt
(\langle\sigma_-(t)\sigma_-(0)\rangle_{\rm{q}}
-\langle\sigma_-\rangle_{\rm{q}}^2),
\end{eqnarray}
which can be again calculated using the quantum regression
theorem \cite{gardiner,Carmichael}. Here $\langle\cdots\rangle_{\rm{q}} $ denotes the average over
the qubit in the quasi-steady state.

To proceed we introduce polar coordinates $r$ and $\varphi$, denoting the amplitude and phase of the radiation field in a coherent state, respectively.
In the stationary limit, we expect that the field distribution function is independent of the phase.
Hence we seek for the solutions $\lim_{t\rightarrow \infty}P(\alpha,t)=P_s(r)$.
In this way  we obtain for the steady-state distribution
\begin{eqnarray}
 P_s(r) = N e^{-\Phi(r)},
 \label{Ps}
\end{eqnarray}
normalized such that $2\pi\int dr\,r\,  P_s(r)=1$.
The expression for $\Phi(r)$ along with further details of the derivation is given in appendix \ref{app:FPE}.

The correlation function of the resonator $\langle a^\dag(t) a(0) \rangle$ can be calculated by
introducing the conditional probability $P(r,\varphi,t/r_0,\varphi_0,0)$ 
for the radiation field to be in the coherent state
$(r,\varphi)$ at time $t$, given that it was in the state $(r_0,\varphi_0)$ at  $t=0$.
In terms of distribution functions, the correlation function is given by \cite{Louisell}
\begin{eqnarray}\label{eq:CF_P}
\langle a^\dag(t) a(0) \rangle
&=&\int dr \; r^2 \int d \varphi \;
d\varphi_0 \; dr_0 \; r^2_0 \; P_{\rm s}(r_0)
             \nonumber \\
 &&  \times e^{i(\varphi_0-\varphi)} \;
  P(r,\varphi,t/r_0,\varphi_0,0)
                  \nonumber \\
 &\equiv&  \int dr\; r^2\; W(r,t),
\end{eqnarray}
where $W(r,t)$  obeys the differential equation 
\begin{equation}\label{eq:W}
\partial W/\partial t = \hat L W.
\end{equation}
The explicit form of the differential operator $\hat L$  can be obtained from the FPE equation \cite{Lax} and it is  given in appendix \ref{app:FPE}.
 To solve Eq.~(\ref{eq:W}) we discretize the amplitude $r$ and diagonalize the
$\hat L$ matrix on the resulting lattice. We can then expand 
$ W(r,t) = \sum_k c_k \chi_k(r) e^{-\lambda_k t}$ in eigenfunctions of $\hat L$,
where $\lambda_k$ is the eigenvalue corresponding to the eigenfunction $\chi_k$,
and the coefficients $c_k$ are determined by the initial condition
$W(r,t=0)=2\pi rP_s(r)$.

Each of the two methods described so far, based on different representations of the density matrix,
has its advantages and limitations. The Fock-state representation used in the direct integration of the Liouville equation is exact but, for practical reasons, can only deal with low photon numbers.
The coherent-state representation and FPE approach provide insight into the distinction between the quantum and classical descriptions and can be employed in the large photon number limit.
But the condition $g/\Gamma_{\varphi} \ll 1$ must be satisfied to allow 
truncating the derivatives to second order.
Hence it fails when the coupling is too strong.

\section{Phase diagram and stationary properties}\label{sec:transition}

In this Section we investigate the stationary properties of single-qubit lasers for different 
values of the qubit-resonator coupling $g$ and detuning $\Delta$.
In  all results presented below  we assume $N_{\rm th}=0$ and
keep the bare qubit inversion fixed at $\tau_0=0.975$.
The study presented here serves a two-fold purpose, on one hand it aims at characterizing the state of the radiation field in the different parameter regimes explored in the experiment \cite{astafiev}, on the other hand it gives us the opportunity to compare the FPE and the DILE methods. 

\subsection{Lasing transition}
\begin{figure}[h]
\centering
\includegraphics[width=0.38\textwidth]{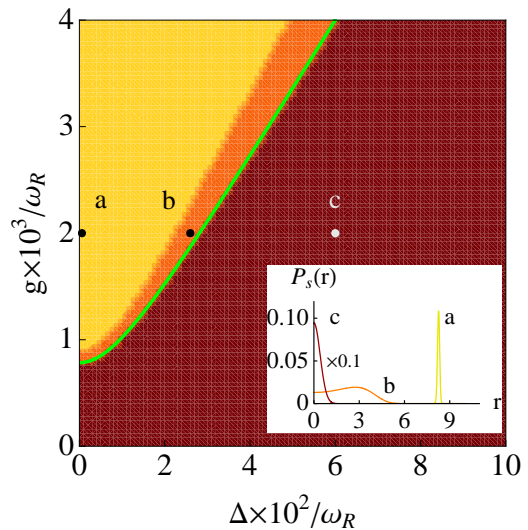}
\caption{(Color online) Phase diagram (and
in the inset distribution functions) in the lasing (light), non-lasing (dark)
and the transition regimes (intermediate) as a function of the 
detuning $\Delta=\omega_{\rm Q} -\omega_{\rm R}$ and coupling $g$ between qubit and resonator.
The distribution functions are plotted with $\kappa = 10^{-4}~\omega_{\rm R}$, $g=0.002~\omega_{\rm R}$
and detuning $\Delta = 0$ at point $a$, $\Delta=0.027~\omega_R$ at point $b$
and $\Delta= 0.06~\omega_{\rm R}$ at point $c$.
Other parameters are $\Gamma_{\varphi}=1.2\times10^{-2}~\omega_{\rm R}$,
$\Gamma_1 = 1.6\times10^{-2}~\omega_{\rm R}$, $\tau_0 = 0.975$ and $N_{\rm th} = 0$.
The solid line represents the transition curve obtained from the semi-classical approximation.
}\label{fig:PD}
\end{figure}

In order to study the nature of the radiation field we first analyze the photon distribution 
function, $P_s(r)$, with results shown in Fig.~\ref{fig:PD}.
For weak coupling or large detuning the system is in the {\sl thermal} regime,
where the resonator behaves like a black-body cavity and $P_s(r)$ decays monotonically as a function of $r$ (curve c in Fig.~\ref{fig:PD}).
Increasing the coupling strength or decreasing the detuning brings the system into 
a  transition regime, where $P_s(r)$ has a maximum at a non-zero value 
of $r$, but the field still has some probability at $r=0$ (curve b).
% The separation between the thermal and the transition regimes can be obtained by minimizing $\Phi(r)$ \cite{Scully}.
A further change of parameters 
pushes the system into the lasing regime with a photon distribution having a negligible weight at $r=0$.
%($P_s(r=0)< 10^{-6}$).
We note that  the lasing threshold (solid line in Fig.\ \ref{fig:PD} ) 
estimated within the semi-classical approximation \cite{brosco2},
\begin{equation}\label{eq:semic}
\frac{2g^2\Gamma_{\varphi}}{\Gamma_{\varphi}^2+\Delta^2} = \frac{\kappa}{\tau_0},
\end{equation}
coincides with the boundary of the thermal regime (found by evaluating whether the minimum of $\Phi (r)$ occurs at $r=0$ \cite{Scully}).

\begin{figure}[h]
\centering
\includegraphics[width=0.42\textwidth]{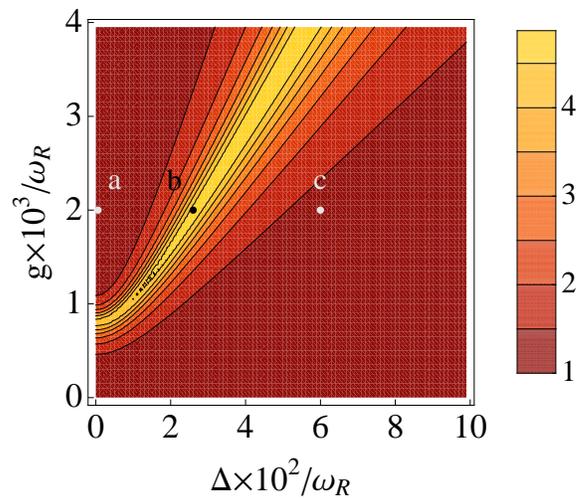}
\caption{(Color online) Fano factor for different values of 
detuning and coupling between qubit and resonator. Points a, b, and c are defined as in Fig.~\ref{fig:PD}.
}\label{fig:fano}
\end{figure}

A further hint at the nature of the radiation field 
in the  different regimes is provided by the Fano factor 
 $F=(\langle n^2\rangle-\langle n\rangle  ^2)/\langle n\rangle$.
As one may expect, deep in the thermal regime the photon distribution is simply a Bose distribution and, since $N_{\rm th}=0$, the Fano factor equals $1$. As one approaches the transition regime, 
due to large fluctuations in the photon number, the Fano factor is strongly enhanced, reaching a
maximum in the transition region. Deep in the lasing regime, the  field is in a coherent state and the Fano factor equals one again, $F\approx 1$.  Fig.~\ref{fig:fano} shows that there is a wide region in the $(\Delta, g)$-plane  where F is significantly larger than $1$. In this region, even well above the semi-classical lasing threshold, the field is not yet  in a coherent state. 

\subsection{Photon number and fluctuations}

\begin{figure}[h]
\centering
\includegraphics[width=0.38\textwidth]{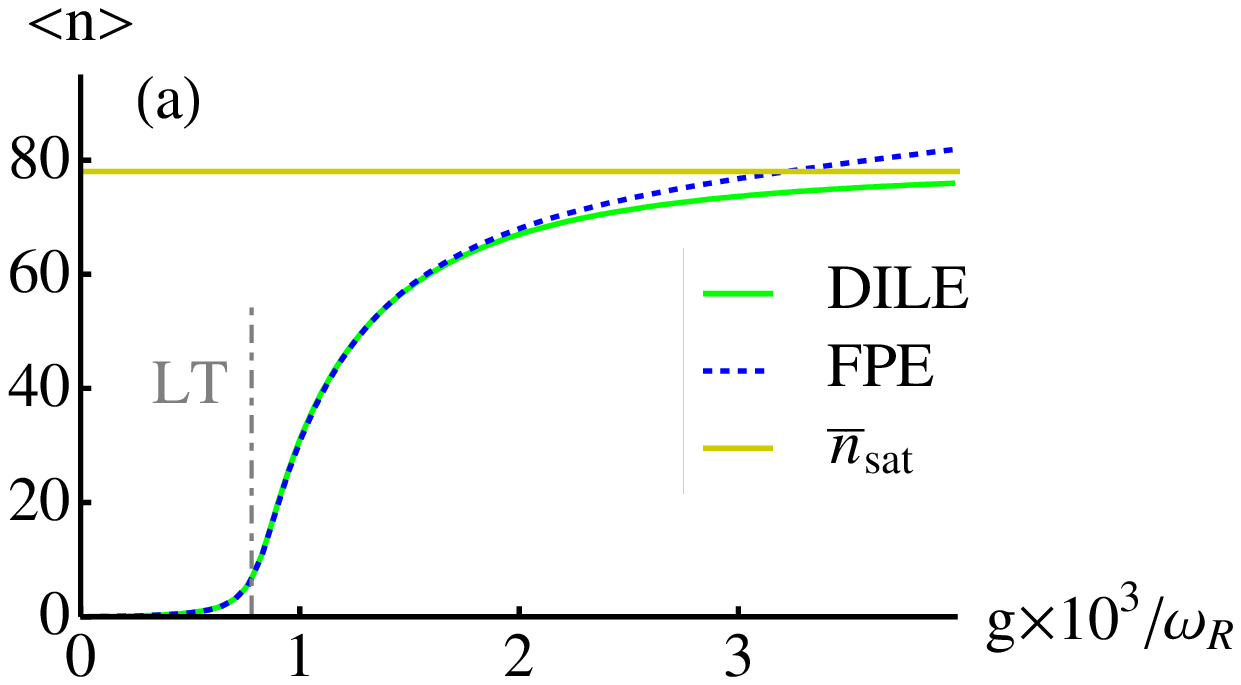}
\includegraphics[width=0.38\textwidth]{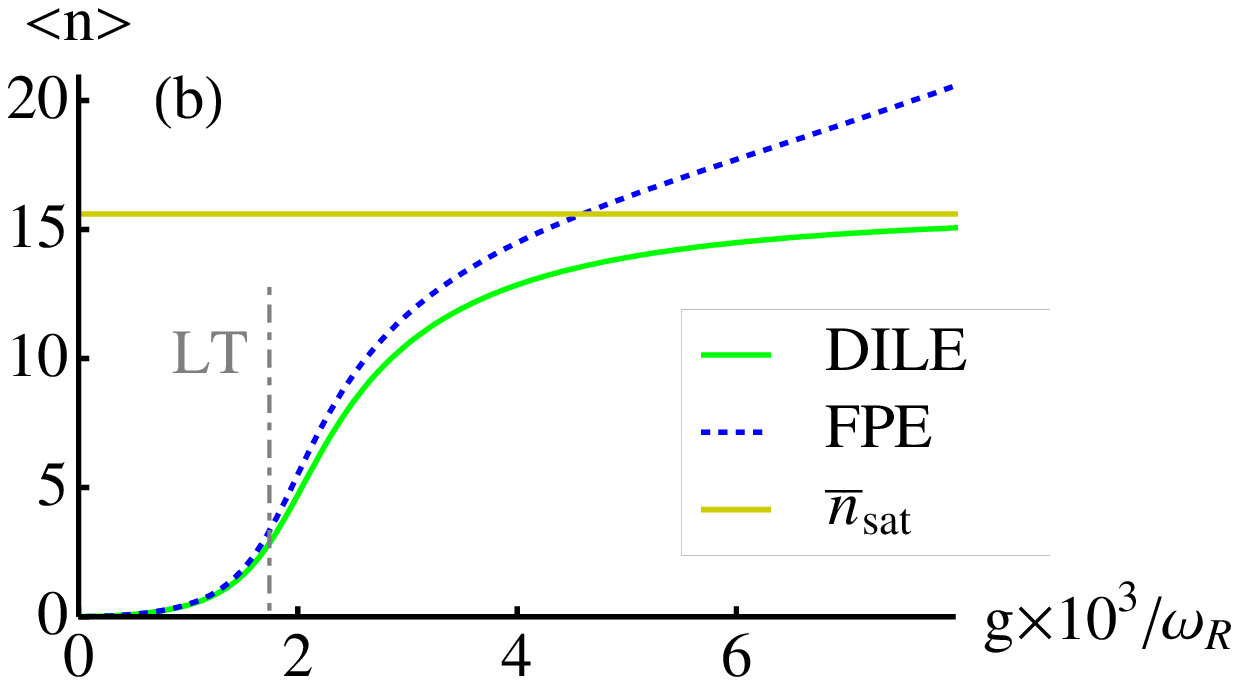}
\caption{(Color online) Average photon number on resonance obtained from the DILE (solid)
and the FPE approach (dotted). Parameters are $\kappa = 10^{-4}~\omega_{\rm R}$ in panel (a)
and $\kappa = 5\times10^{-4}~\omega_{\rm R}$ in panel (b).
The dot-dashed line indicates the lasing threshold (LT) estimated using 
Eq.~(\ref{eq:semic}).
}\label{fig:nf}
\end{figure}

The lasing transition is evident in the average photon number
$\langle n\rangle$. Results are shown
 in Fig.~\ref{fig:nf} for varying coupling strength $g$. In the
non-lasing regime, the photon number is small. 
Remarkably, even for a single-qubit laser $\langle n\rangle$ increases
sharply near the threshold before it saturates deep in the  lasing
regime. Within the semi-classical approximation, the saturation
photon number can be estimated as
\begin{equation}
 \bar{n}_{\rm{sat}}=\frac{\Gamma_1\tau_0}{2\kappa}.
\end{equation}

In Fig.\ \ref{fig:nf} we compare the results obtained
from the DILE and FPE methods.
Both coincide for weak coupling $g$. They differ 
when the coupling becomes stronger since the
condition $g/\Gamma_\varphi\ll 1$ needed for the FPE approach is violated.
The comparison of the two panels in  Fig.\ \ref{fig:nf} shows that both approaches describe the lasing transition well if the bare damping rate of the resonator is weak (upper panel), but for a higher damping rate (lower panel) the FPE approach is not sufficient.

\begin{figure}[h]
\centering
\includegraphics[width=0.38\textwidth]{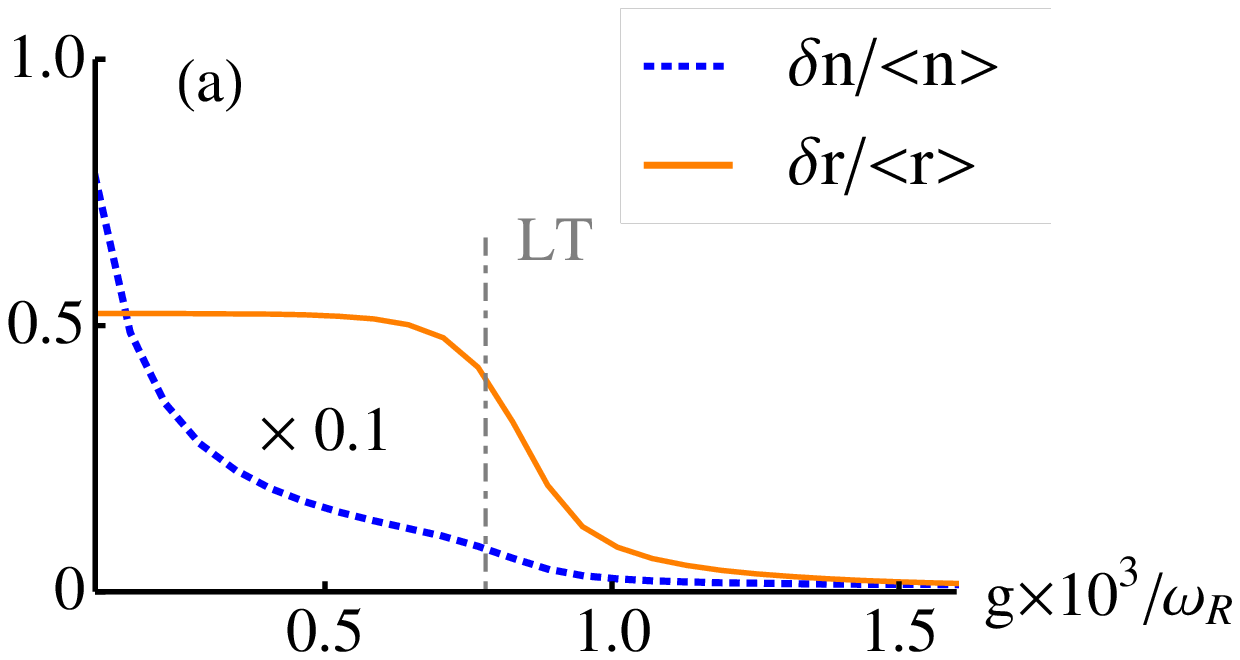}
\includegraphics[width=0.38\textwidth]{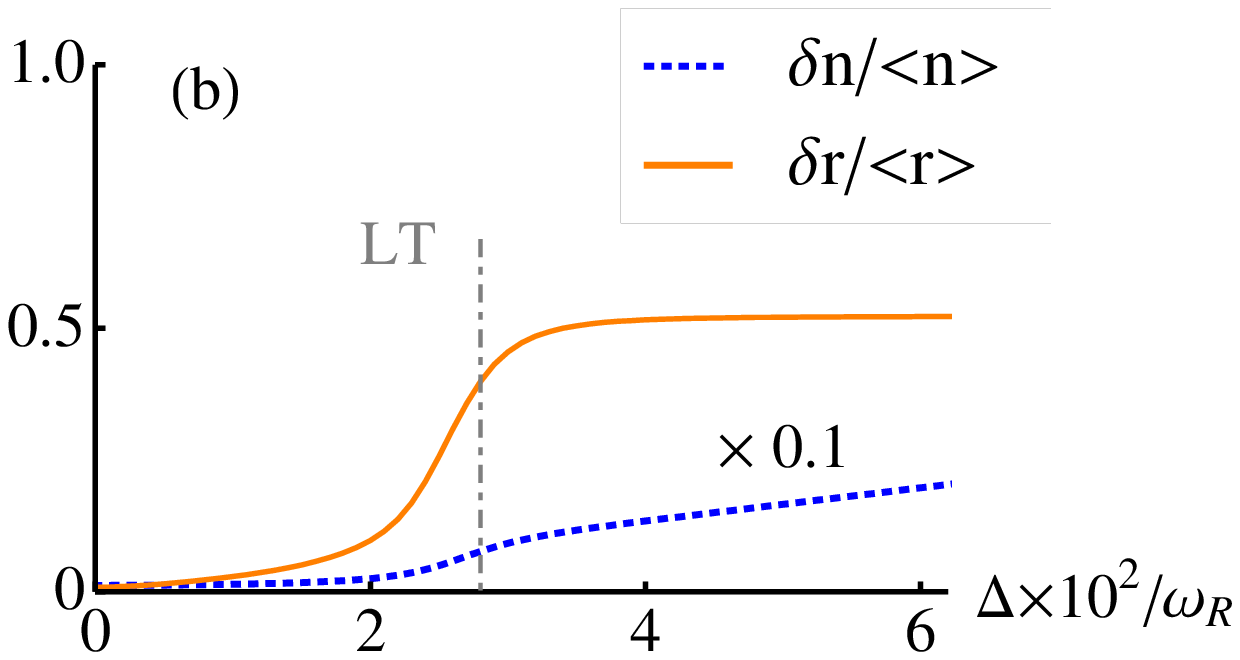}
\caption{(Color online) Relative fluctuations as function of $g$ at resonance (panel a)
and as function of $\Delta$ at $g = 0.002~\omega_{\rm R}$ (panel b).
In both panels $\kappa = 10^{-4}~\omega_{\rm R}$.
}\label{fig:fl}
\end{figure}

Of interest for the following discussions are also the relative fluctuations of the photon number
$\delta n/\langle n\rangle$ and of the field amplitude $\delta r/\langle r\rangle$ shown in Fig.\ \ref{fig:fl}. (For the parameters used, DILE and FPE coincide well.)
The relative fluctuations, although they are rather weak in the lasing regime,
have important effects on the spectral properties of the single-qubit laser, which we will discuss in the next section.
%: The amplitude fluctuations are coupled to the phase fluctuations, thus giving rise to a large contribution to the laser linewidth and resulting in the breakdown of the phase diffusion model.

In the non-lasing regime, the correlations between the qubit and the resonator are negligible
and the distribution function for the steady state can be approximated as
$P_{\rm s} (r) = \pi e^{-b\;r^2}/b$ with
\begin{eqnarray}
 b = \frac{\tau_0}{2(1+\tau_0)} \left(1-\frac{2\tau_0\Gamma_\varphi}{\kappa}\frac{g^2}{\Gamma_\varphi^2+\Delta^2}\right),
\end{eqnarray}
which is positive in this regime.
The relative fluctuations of the photon number are thus given by $\delta n/\langle n\rangle = \sqrt{1+b}$,
which grow with decreasing $g$ or increasing $\Delta$.
The relative amplitude fluctuations are constant in this regime, namely,
$ \delta r/\langle r\rangle = \sqrt{4/\pi-1} \simeq 0.5$, as shown in Fig. \ref{fig:fl}.

\section{Laser linewidth }\label{sec:linewidth}
This section is devoted to the study of the emission spectrum of the single-qubit laser 
radiation and the dependence of its linewidth on the qubit-resonator coupling and detuning. 
Both in the FPE and the DILE approaches the correlation function of the resonator can be expanded 
as
$\langle a^{\dagger}(t) a(0) \rangle = \sum_k \alpha_k \exp{(-\lambda_k t)}$,
where $-\lambda_k$ are the eigenvalues of the discretized differential operator $\hat L$ 
defined in Appendix \ref{app:FPE}
 or of the superoperator $G$ introduced in Section \ref{subsec:DirectInt} (both in the rotating frame), respectively.
Using this expansion in Eq.~\eqref{eq:Spectrum-Def} we obtain
\begin{eqnarray}
  && S_{\rm e}(\omega) = 2 \sum_k \frac{1}
 {[\omega+\textrm{Im}(\lambda_k)]^2+[\textrm{Re}(\lambda_k)]^2}
                     \nonumber \\
&& \times\left\{\textrm{Re}(\alpha_k)\textrm{Re}(\lambda_k)
 +\textrm{Im}(\alpha_k)[\omega+\textrm{Im}(\lambda_k)]
 \right\} \, . \label{eq:Spectrum}
\end{eqnarray}
This equation serves as starting point for the calculation of
the resonator spectrum once the eigenvalues $\lambda_k$ and the coefficients $\alpha_k$ are known.

\begin{figure}[h]
\centering
\includegraphics[width=0.4\textwidth]{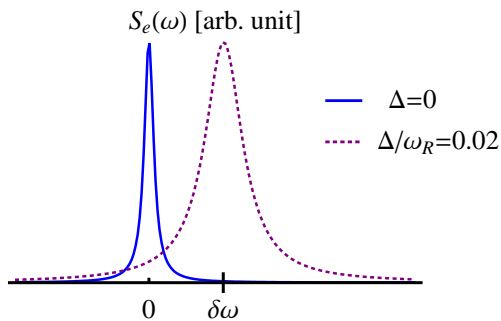}
\caption{(Color online) Emission spectrum $S_{\rm e}(\omega)$ in the rotating frame
for $g=0.003~\omega_{\rm R}$ and $\kappa = 5\times10^{-4}~\omega_{\rm R}$.
The frequency shift $\delta \omega$ at $\Delta = 0.02~\omega_{\rm R}$ is about $ 0.56~\kappa$.
}\label{fig:Spectrum}
\end{figure}

Results for two different values of the detuning are plotted in Fig.~\ref{fig:Spectrum}. 
The spectrum is characterized by a linewidth, which we define -  even in cases where the spectrum is not 
 Lorentzian - to be the half-width at half-maximum (HWHM). We note that a detuning between qubit and resonator shifts the emission spectrum away from the bare frequency $\omega_{\rm R}$ of the resonator. Near resonance, this frequency shift grows linearly with the detuning $
\Delta$ \cite{brosco1,brosco2}.

When analyzing the emission spectrum using the two approaches described above
we arrive at the following main conclusions:

(i)  The linewidth depends in a non-monotonous way
on the coupling strength, consistent with Refs.~\onlinecite{brosco1,brosco2}. 

(ii) Away from resonance (but deep in the lasing regime), amplitude fluctuations
contribute significantly to the linewidth. But they are neglected in the phase diffusion model.

(iii) For strong coupling or weak damping of the resonator the linewidth is strongly enhanced
in the transition regime. In this case the emission spectrum is no longer purely Lorentzian.

\subsection{Dependence on the coupling strength}

\begin{figure}[h]
\centering
\includegraphics[width=0.42\textwidth]{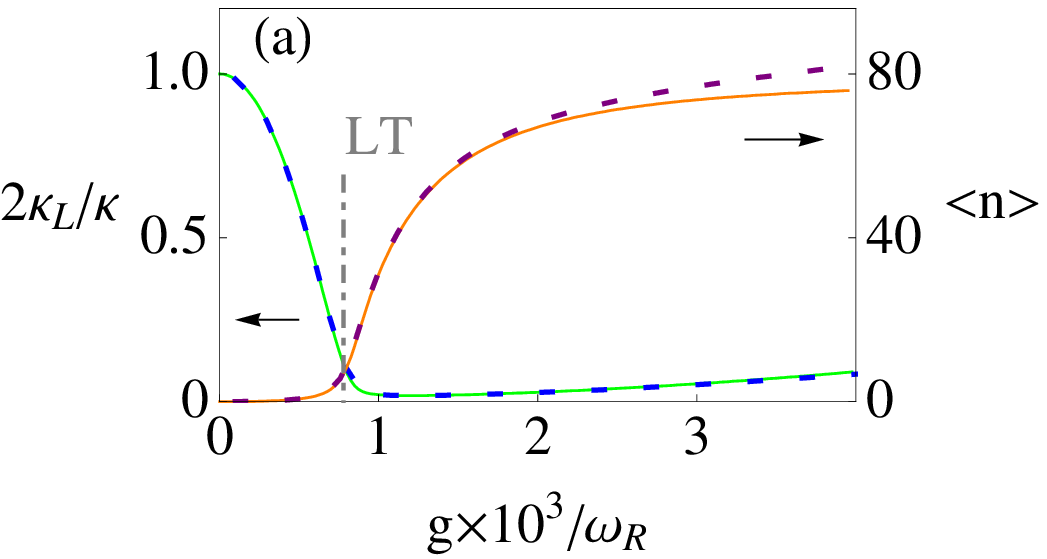} \\[3mm]
\includegraphics[width=0.42\textwidth]{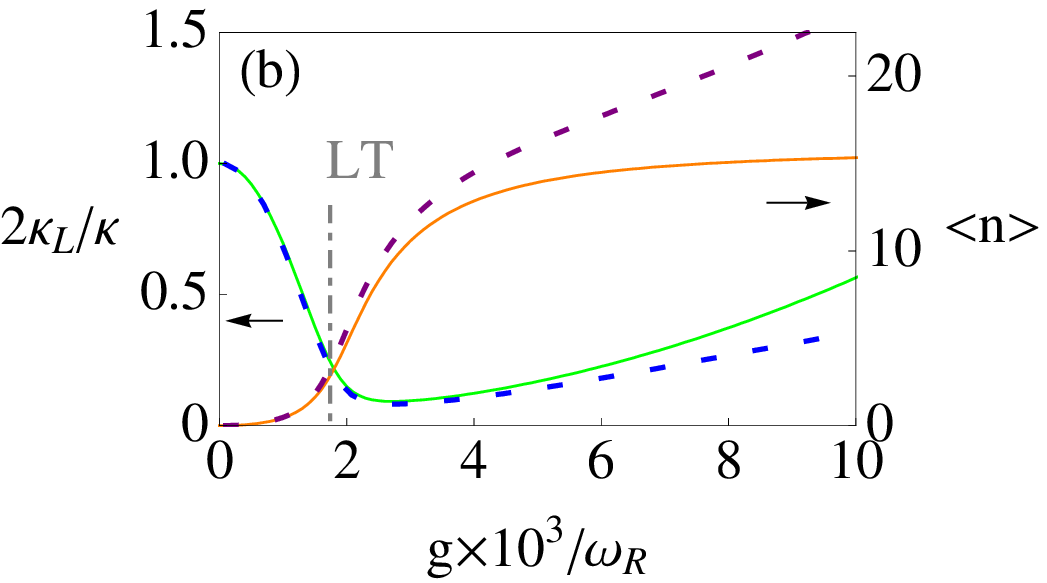}
\caption{(Color online) Average photon number and linewidth as functions of the coupling strength on resonance
with $\kappa = 10^{-4}~\omega_{\rm R}$ in panel (a) and $\kappa = 5\times10^{-4}~\omega_{\rm R}$ in panel (b). Results obtained from
the DILE and the FPE methods are represented by the solid and the dashed lines, respectively.
} \label{fig:lwvsg}
\end{figure}

One of the central results of Refs.\ \onlinecite{brosco1,brosco2} 
was the non-monotonic dependence of the linewidth on the qubit-resonator coupling strength.
This is demonstrated in Fig.\ \ref{fig:lwvsg}, where the average photon number in the resonator and the linewidth of the emission spectrum are plotted as functions of the coupling strength $g$ at resonance, $\Delta=0$.
We see that, while the photon number rapidly increases at the lasing transition and then saturates, the linewidth shows a non-monotonic behavior. For increasing but still weak coupling we observe the linewidth narrowing which is typical for lasers. However, in the deep lasing regime, the linewidth grows again with coupling strength, leading to a deterioration of the lasing state. This effect is very pronounced for low photon numbers in the resonator and  strong coupling. In this regime, the linewidth can even become larger than the bare linewidth of the resonator.

As described in Sec.~\ref{sec:model}, the DILE can be used if the photon number is not too high, i.e., if the damping rate $\kappa$ of the resonator is not too small.
On the other hand, the FPE approach requires that the coupling strength is much smaller than the dephasing rate of the qubit, $g \ll \Gamma_{\varphi}$.
Figs.~(\ref{fig:nf}) and (\ref{fig:lwvsg}) show good agreement between both approaches for parameters where both are valid, i.e., for weak coupling and low photon number.

\subsection{Linewidth as function of the detuning}

At resonance (deep in the lasing regime), a satisfactory picture of the linewidth properties
can be derived using the phase diffusion model (PDM) \cite{Scully,haken}.
Within this model the effects of amplitude fluctuations on the spectrum 
(see below) are neglected, and  the linewidth coincides with the phase diffusion rate, 
which is only affected by phase fluctuations.

\begin{figure}[h]
\centering
\includegraphics[width=0.42\textwidth]{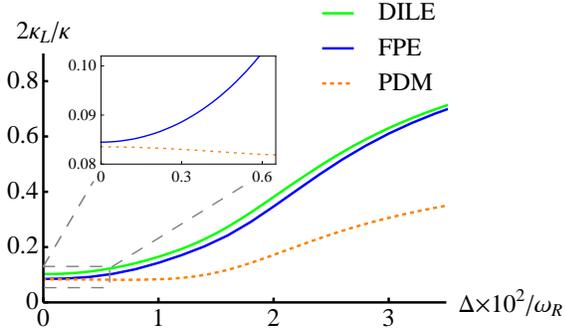}
\caption{(Color online) Linewidth as a function of the detuning with 
$\kappa = 5\times10^{-4}\omega_{\rm R} $,
$g=0.003~ \omega_{\rm R}$. }\label{fig:lwd}
\end{figure}

The PDM turns out to be no longer valid away 
from resonance when the quantum noise dominates.
As shown in Fig.~\ref{fig:lwd}, when amplitude fluctuations are taken into account,
the laser linewidth \emph{increases} with growing detuning.
This is obtained both in the DILE and the FPE approaches and 
in agreement with findings of Ref.~\onlinecite{didier}.
In contrast the PDM, although working  well at resonance,
predicts the wrong slope and curvature of the linewidth as a function of detuning.

The failure of the PDM is due to the coupling of phase and amplitude fluctuations.
In order to describe their effects on the linewidth, we consider the FPE in polar coordinates
(see appendix \ref{app:FPE} for details),
\begin{eqnarray}\label{eq:FPEinPC}
\frac{\partial P(r,\varphi,t)}{\partial t} &=&
 \frac{\kappa}{4r}\frac{\partial}{\partial r}
 \Bigl\{ [ N_{\rm th}+Q(r)+2u(r)]r\frac{\partial}{\partial r}
              \nonumber \\[1mm]
&+&4v(r)\frac{\partial}{\partial\varphi } +F(r)
\Bigr\}P(r,\varphi,t)
             \nonumber \\
&\!+\!& \!\kappa\Bigl( D_2(r) \frac{\partial^2}{\partial \varphi^2}
+D_1(r) \frac{\partial}{\partial \varphi} \Bigr) P(r,\varphi,t) \, .
             \nonumber \\
\end{eqnarray}
The mixed derivative 
$(\propto \partial^2/\partial r \partial\varphi)$
couples the dynamics of phase and amplitude. The coefficient $v(r)$
 is given by
\begin{eqnarray}
 v(r) = \textrm{Im} \left[e^{-2i\varphi} R(r,\varphi)\right].
\end{eqnarray}
In the limit of full population inversion, $\tau_0 =1$, 
and vanishing pure dephasing, $\Gamma_\varphi^* =0$,
it reduces to
\begin{eqnarray}
 v(r) = \frac{\Delta\, r^2\left[(5+\Delta^2/\Gamma_\varphi^2)n_0+2 r^2\right]}
 {4\kappa\left[(1+\Delta^2/\Gamma_\varphi^2)n_0+r^2\right]^3}.
 \label{v(r)}
\end{eqnarray}
This form illustrates that the coupling term vanishes on resonance.
But off-resonance in the lasing regime, $v(r)$ couples amplitude fluctuations 
to the phase dynamics and - as shown below - qualitatively influences the linewidth.

\begin{figure}[h]
\centering
\includegraphics[width=0.4\textwidth]{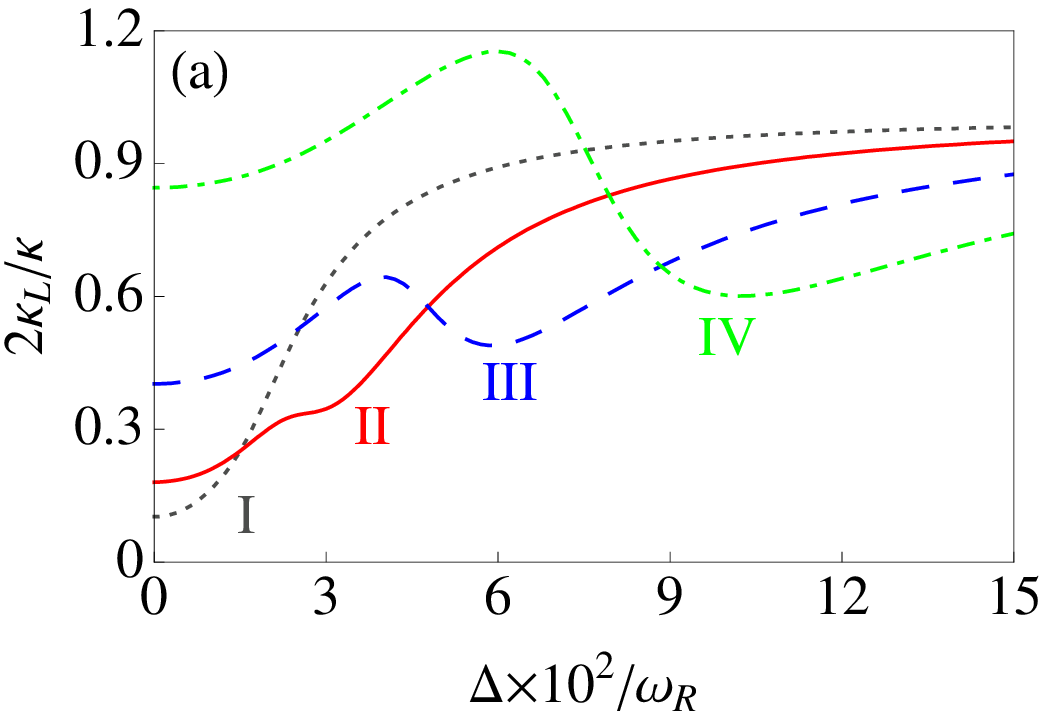}\\[3mm]
\includegraphics[width=0.4\textwidth]{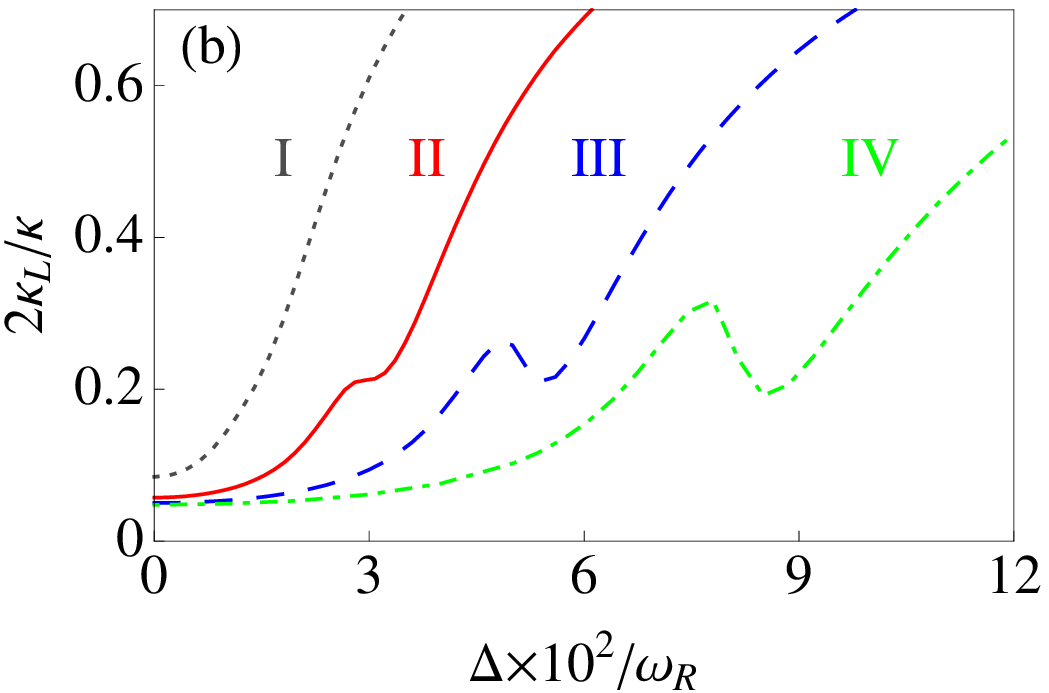}
\caption{(Color online) Linewidth as functions of the detuning obtained from the DILE approach in panel (a)
and from the FPE method in panel (b). Parameters are given in Table. \ref{tab:para}. }
\label{fig:lwlgsk}
\end{figure}

In the transition regime, the linewidth is enhanced and,
as shown in Fig.\ \ref{fig:lwlgsk}, exhibits a peak structure for 
strong coupling or low damping of the resonator.
This phenomenon arises due to the coupling of amplitude and phase fluctuations. 
For an estimate we expand $v(r)$ of Eq.~(\ref{v(r)})
(neglecting pure dephasing and setting $\tau_0=1$) near the 
 lasing transition in the small amplitude of the radiation field,  $r^2 \ll \bar{n}_{\rm{sat}}$.
The result is $v(r) \simeq a_1 r^2/\bar{n}_{\rm{sat}}$ with expansion coefficient
\begin{eqnarray}
a_1 =  \frac{\sqrt{4 \xi -1}}{4 \xi} (1+\xi).
\end{eqnarray}
In the derivation of the above equation, we use the relation between
the coupling strength $g$, the damping rate of the resonator $\kappa$,
and the detuning $\Delta$ at the transition, which is given by Eq.\ (\ref{eq:semic}).
The expansion coefficient $a_1$ depends on the ratio between coupling and damping rate,
\begin{eqnarray}
\xi =  \frac{g^2}{\kappa\Gamma_1}.
\end{eqnarray}
It grows with increasing $g$ or decreasing $\kappa$,
which leads to an increased coupling of the amplitude fluctuations and hence increased linewidth.

In the two panels of Fig.\ \ref{fig:lwlgsk} we plot the linewidth as a function of the detuning for different $g$ and $\kappa$, as obtained in the two approaches.
Since the coefficient $a_1$ depends on the ratio $\xi$,
we choose the parameters such that each curve in Fig.\ \ref{fig:lwlgsk} (a) has a corresponding one
in Fig.\ \ref{fig:lwlgsk} (b) which has the same value for $\xi$, as shown in Table \ref{tab:para}.
The linewidths represented by curves with the same ratio of coupling to damping rate show similar behavior around the lasing transition.
When $\xi$ is small (\textit{e.g.} $\xi = 1.125$), the linewidth grows monotonically with detuning and
no peculiar features show up. For larger values of $\xi$, the linewidth is enhanced around the transition regime, which becomes more pronounced
as $\xi$ increases. Moreover, for strong coupling, e.g., for $g/\omega_{\rm R}=0.012$, the peak of the linewidth can be even exceed the bare value, $\kappa/2$. 
In this case, the linewidth at resonance $\kappa_{\rm{L}}(\Delta=0)$ is already comparable to $\kappa/2$.

\begin{table}
 \begin{tabular}{|l|c|c|}
  \hline
                               & ~Fig.\ \ref{fig:lwlgsk} (a) :  & ~Fig.\ \ref{fig:lwlgsk} (b) :             \\
                               & ~$\kappa = 5\times10^{-4} $    & ~$g = 0.003$                     \\[1mm] \hline
  ~curve I: $\xi = 1.125$     & ~$g = 0.003$                   & ~$\kappa = 5.0\times10^{-4}$      \\[1mm] \hline
  ~curve II: $\xi = 3.125$    & ~$g = 0.005$                   & ~$\kappa = 1.8\times10^{-4}$      \\[1mm] \hline
  ~curve III: $\xi = 8$       & ~$g = 0.008$                   & ~$\kappa = 7.2\times10^{-5}$      \\[1mm] \hline
  ~curve IV: $\xi = 18$       & ~$g = 0.012$                   & ~$\kappa = 3.1\times10^{-5}$      \\[1mm] \hline
 \end{tabular}
 \caption{ Parameters in Fig.\ \ref{fig:lwlgsk}. Here $g$ and $\kappa$ are in unit of $\omega_{\rm R}$.}
 \label{tab:para}
\end{table}

In the non-lasing regime for strong detuning, the linewidth can be well approximated by \cite{brosco2}
\begin{eqnarray}
\kappa_{\rm{L}} = \frac{\kappa}{2} - \frac{g^2\Gamma_{\varphi}}{\Gamma_{\varphi}^2+\Delta^2} \tau_0 < \frac{\kappa}{2},
\end{eqnarray}
approaching the bare linewidth of the resonator, $\kappa/2$, in the infinite detuning limit.

\subsection{Non-exponential decay}

\begin{figure}[h]
\centering
\includegraphics[width=0.4\textwidth]{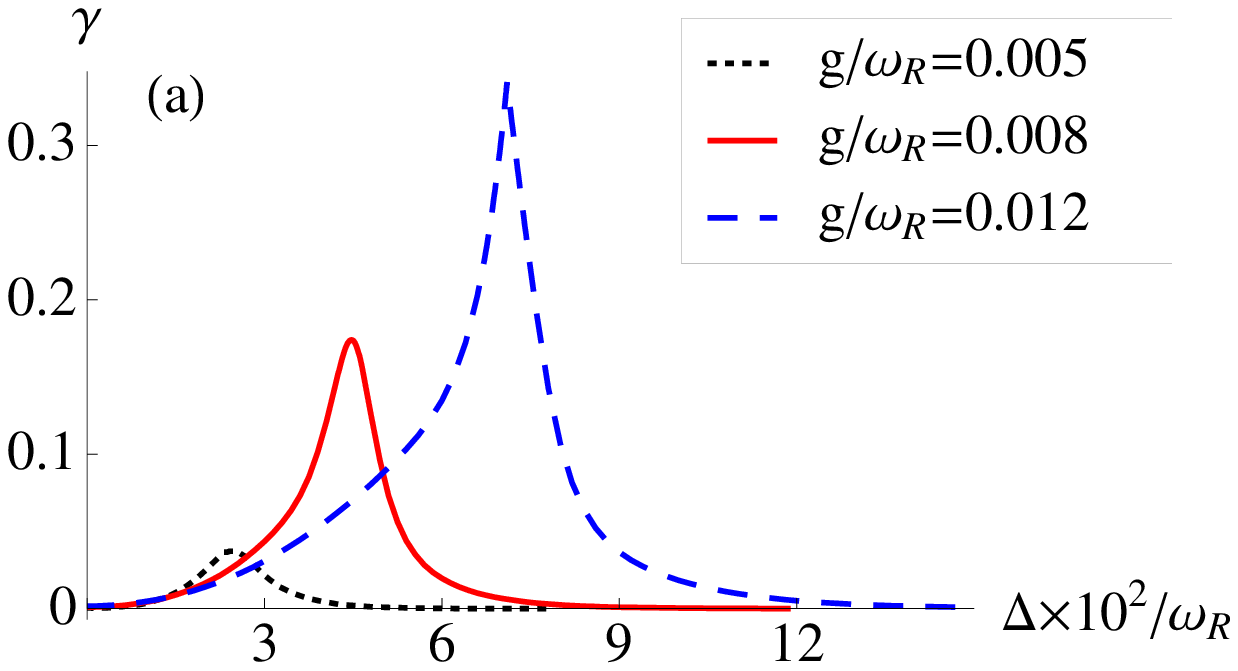}\vspace{7mm}
\includegraphics[width=0.4\textwidth]{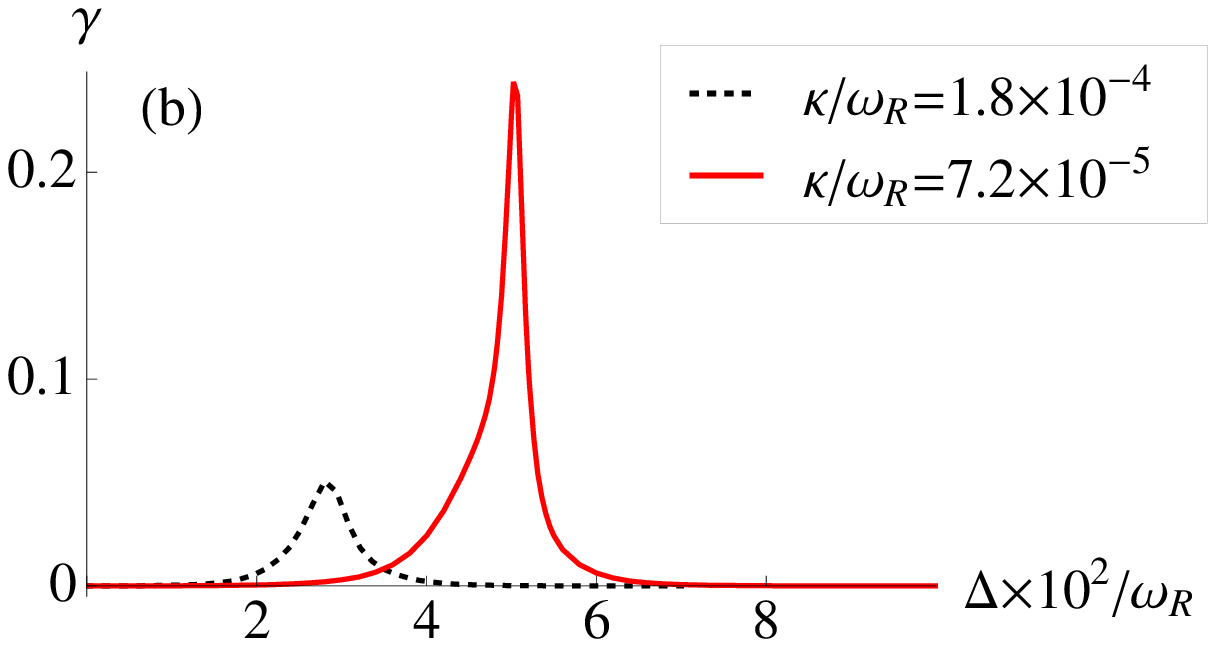}
\caption{(Color online) Relative distance $\gamma$ between the 
HWHM and the real part of the first eigenvalue
as a function of the detuning for different values of $g$ with 
$\kappa/\omega_{\rm R} = 5\times10^{-4}$ (DILE)
in panel (a) and for different values of $\kappa$ with 
$g/\omega_{\rm R} = 0.003$ (FPE) in panel (b).
}\label{fig:Decay}
\end{figure}
Generally, as described in Eq.\ (\ref{eq:Spectrum}), 
the emission spectrum $S_{\rm e}(\omega)$ is the sum of contributions
from different eigenvalues.
We find that both in the deep lasing and the non-lasing regimes, 
a single eigenfunction, the one corresponding
to the eigenvalue with the smallest real part (denoted as $\lambda_1$), is sufficient to describe the spectral properties.
In this case, the correlation function of the radiation field decays exponentially in time 
and the spectrum is a  Lorentzian.
However, in the transition regime, where the linewidth shows the non-monotonic behavior, 
more than one eigenfunction contributes to the spectrum.

As a measure of the deviation of the correlation function from the simple exponential form,
we define the relative distance between the HWHM 
linewidth $\kappa_{\rm{L}}$ and the real part of $\lambda_1$,
\begin{eqnarray}
 \gamma =  \frac{\left|\kappa_{\rm{L}}-\textrm{Re}\left[\lambda_1\right]\right|}{\kappa_{\rm{L}}}.
\end{eqnarray}
Fig.~\ref{fig:Decay} shows $\gamma$ as a function of detuning for different values of $g$ and $\kappa$.
At resonance (deep lasing regime) and  far off-resonance (non-lasing regime)
the first eigenvalue is sufficient to describe the linewidth, and the
decay of the correlation function is simply exponential.
However, when approaching the transition regime where a peak in the linewidth shows up, 
the weights of further eigenvalues increase, indicating that more eigenvalues must be taken into account.
More details are presented in Appendix \ref{app:eigenvalues}.

\section{Conclusion}\label{sec:discussion}

We have studied a single-qubit laser consisting of a qubit
coupled strongly to the resonator. The lasing behavior was investigated in
a parameter space spanned by the coupling strength $g$ and the detuning $\Delta$.
Increasing $g$ or decreasing $\Delta$ pushes the system towards the lasing state,
characterized by an increase of the average photon number and a decrease of its relative fluctuations.

In single-qubit lasers at low temperature the linewidth of the emission spectrum
is dominated by the quantum noise arising from the correlation between the qubit and the resonator.
On resonance, at $\Delta=0$, the linewidth can be described by the phase diffusion model (PDM)
for the phase fluctuations of the laser field.
In this case, the emission spectrum is Lorentzian with linewidth simply given by the diffusion coefficient in the Fokker-Planck equation.
Away from resonance but still deep in the lasing regime, 
although the amplitude fluctuations are still weak,
their contribution to the linewidth is magnified via the coupling to the phase fluctuations.
In this case, the PDM is no longer sufficient.

With increasing coupling strength $g$ or decreasing damping rate of the resonator, 
the coupling between the phase and amplitude dynamics and, hence,
 the contribution of amplitude fluctuations to the linewidth becomes stronger.
In the transition regime to the lasing state the correlation function of the resonator does not 
simply decay exponentially in time,
and the Fourier transform is no longer Lorentzian.
Their behavior is no longer governed by a single eigenvalue $\lambda_1$, 
 distinguished from the other ones by having a much smaller real part. 
Instead several eigenvalues, with similar real parts, contribute to the spectrum.

In our studies, we employed two complementary approaches in order to
cover a wide range of parameters.
The DILE method in the Fock state representation is free from approximations on the
relative strengths of the coupling, damping rate of the resonator, and that of the qubit.
A practical limitation arises from the fact that it can deal only with low photon numbers. 
In the large-photon number limit,
the FPE approach can be employed. The constraint on it is set by the approximation
that the coupling strength should be  lower than the damping rate of the qubit.
In an overlapping parameter regime, where both methods are sufficient, results obtained in the two approaches show good agreement.

\section*{Acknowledgements}
The authors wish to acknowledge helpful discussions with M.~Marthaler
and N.~Didier. PQJ was supported by the Baden-W\"urttemberg Stiftung 
via the Kompetenznetz Funktionelle Nanostrukturen.
AR acknowledges support by the Humboldt foundation.

\vspace{5mm}
\appendix

\section{FPE in polar coordinates}\label{app:FPE}

In polar coordinates, the FPE is given by
\begin{eqnarray}
\frac{\partial P(r,\varphi,t)}{\partial t} &=&
 \frac{\kappa}{4r}\frac{\partial}{\partial r}
 \Bigl\{ [ N_{\rm th}+Q(r)+2u(r)]r\frac{\partial}{\partial r}
              \nonumber \\[1mm]
&+&4v(r)\frac{\partial}{\partial\varphi } +F(r)
\Bigr\}P(r,\varphi,t)
             \nonumber \\
&\!+\!& \!\kappa\Bigl( D_2(r) \frac{\partial^2}{\partial \varphi^2}
 +D_1(r) \frac{\partial}{\partial \varphi} \Bigr) P(r,\varphi,t),
             \nonumber \\
\end{eqnarray}
where $u(r)$ and $v(r)$ denote the real and imaginary parts of $e^{-2i\varphi} R$, respectively,
and
\begin{eqnarray}
F(r)&=& r\frac{d}{d r}\Bigl[ N_{\rm th}+Q(r)+2u(r)\Bigr]
             \nonumber \\
 &+&2 r^2 \Bigl[1-\frac{C_0}{X(r)}\Bigr]+4u(r),
             \nonumber \\
D_2(r) &=&\frac{N_{\rm th}+Q(r)-2u(r)}{4 r^2} ,
             \nonumber \\
D_1(r) &=& \frac{\Delta C_0 }{2\Gamma_\varphi X(r)}
+\frac{v(r)}{r^2},
             \nonumber \\[2mm]
X(r) &=& \frac{n_0(\Gamma_\varphi^2+\Delta^2)}{\Gamma_\varphi^2(n_0+r^2)}.
\end{eqnarray}
In the steady state we expect the distribution function not to depend
on the phase. It  can be written  in the form of Eq.~(\ref{Ps})
with
\begin{eqnarray}
 \Phi(r) &=& \ln \left[ N_{\rm th}+Q(r)+2u(r)\right]
              \nonumber \\[1mm]
&& + 2\!\int \!dr  \, \frac{ r^2 \left[1-C_0/X(r) \right] \!+\! 2u(r)}
 {r \left[ N_{\rm th}+Q(r)+2u(r)\right]}.
\end{eqnarray}

The correlation function of the resonator $\langle a^\dag(t) a(0) \rangle$ can be obtained by
introducing  \cite{Lax},
\begin{eqnarray}
 W(r,t) &=& \int d \varphi \;
d\varphi_0 \; dr_0 \; r^2_0 \; P_{\rm s}(r_0)
             \nonumber \\
 &&  \times e^{i(\varphi_0-\varphi)} \;
  P(r,\varphi,t/r_0,\varphi_0,0),
\end{eqnarray}
which obeys the differential equation
\begin{eqnarray}
\frac{\partial W}{\partial t} &\equiv& \hat L W
             \nonumber \\
& =& \frac{\kappa}{4r}\frac{\partial}{\partial r}
 \Bigl\{ [ N_{\rm th}+Q(r)+2u(r)]r\frac{\partial}{\partial r}
              \nonumber \\[1mm]
&+& 4 i v(r) + F(r)
\Bigr\}W
             \nonumber \\
&\!+\!& \!\kappa\Bigl( -D_2(r)
 + i D_1(r) \Bigr) W. \\ \nonumber
\end{eqnarray}
We numerically solve this problem by discretizing the amplitude $r$ and diagonalize the
$\hat L$ matrix on the lattice. We expand $W(r,t)$ in terms of eigenfunctions of $\hat L$, namely,
$ W(r,t) = \sum_n c_n \chi_n(r) e^{\lambda_n t}$ with
$\lambda_n$ being the eigenvalue corresponding to the eigenfunction $\chi_n$.
The expansion coefficients are determined by the initial condition
$W(r,t=0)=2\pi rP_{\rm s}(r)$.

\section{Eigenvalues contributing to the linewidth}\label{app:eigenvalues}

\begin{figure}[t]
\centering
\includegraphics[width=0.4\textwidth]{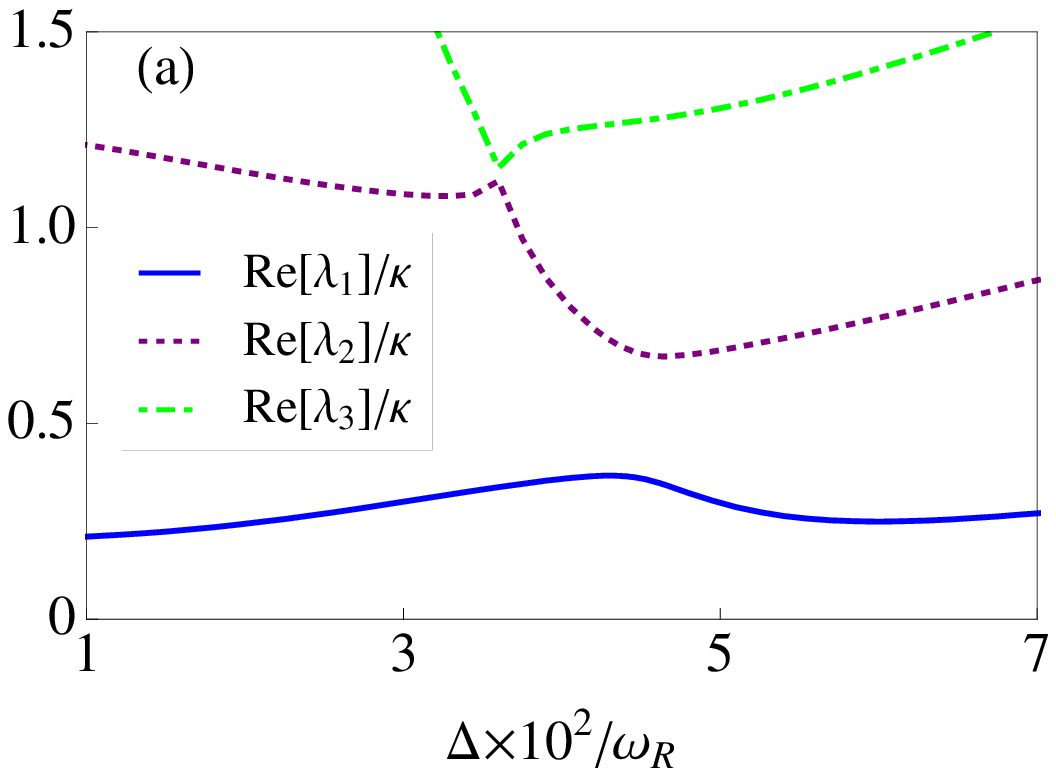}\\[5mm]
\includegraphics[width=0.4\textwidth]{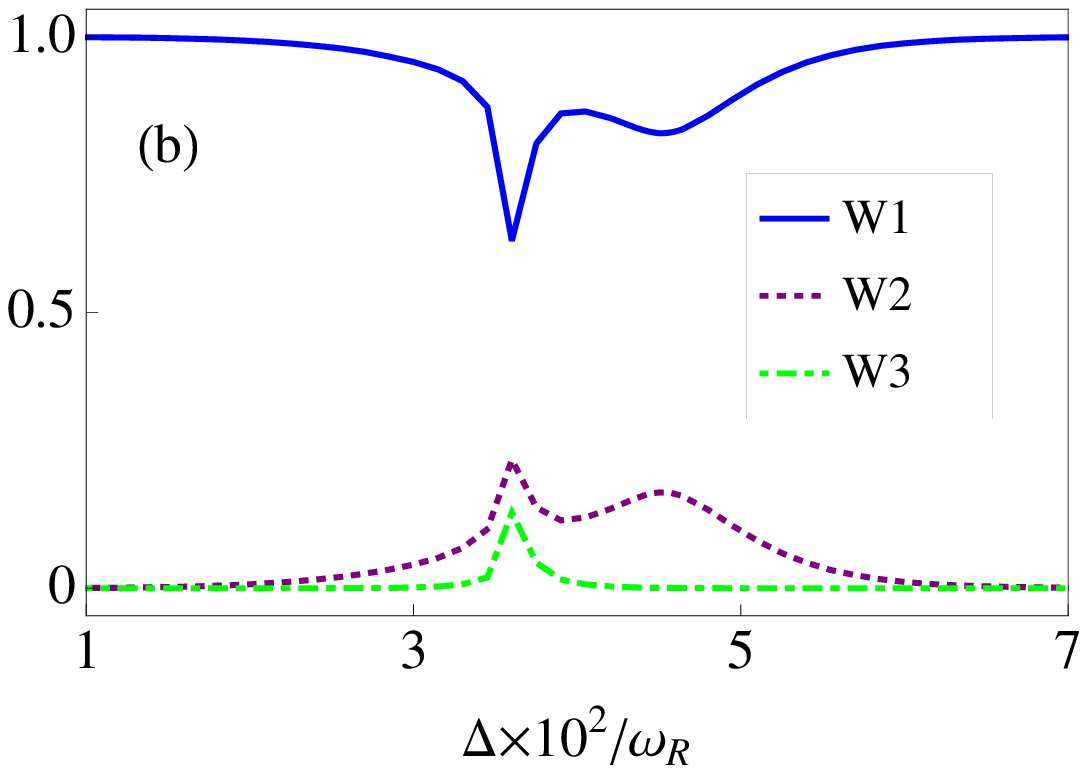}
\caption{Real parts of the eigenvalues $\lambda_i$ (panel (a)) and their weights $W_i$
(panel (b)) as function of the detuning. The results are obtained from the DILE method
with $\kappa = 5\times10^{-4}~\omega_{\rm R}$ and $g = 0.008~\omega_{\rm R}$.
}\label{fig:eigeng}
\end{figure}

\begin{figure}[t]
\centering
\includegraphics[width=0.4\textwidth]{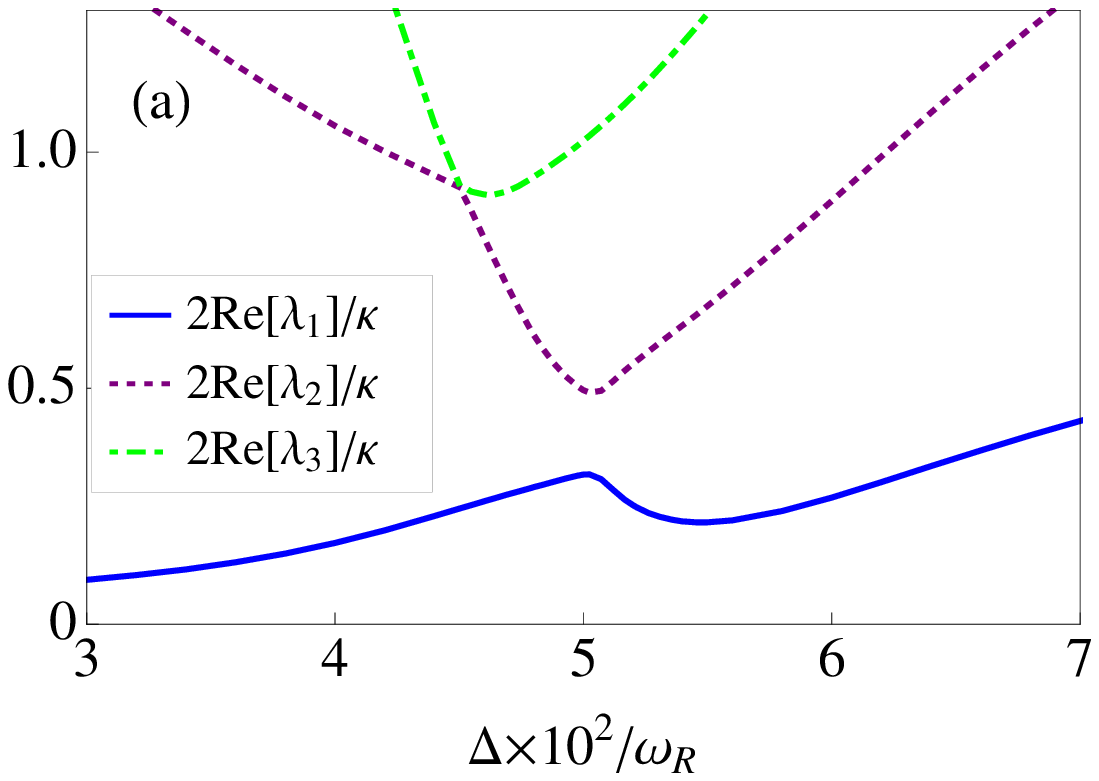}\\[5mm]
\includegraphics[width=0.4\textwidth]{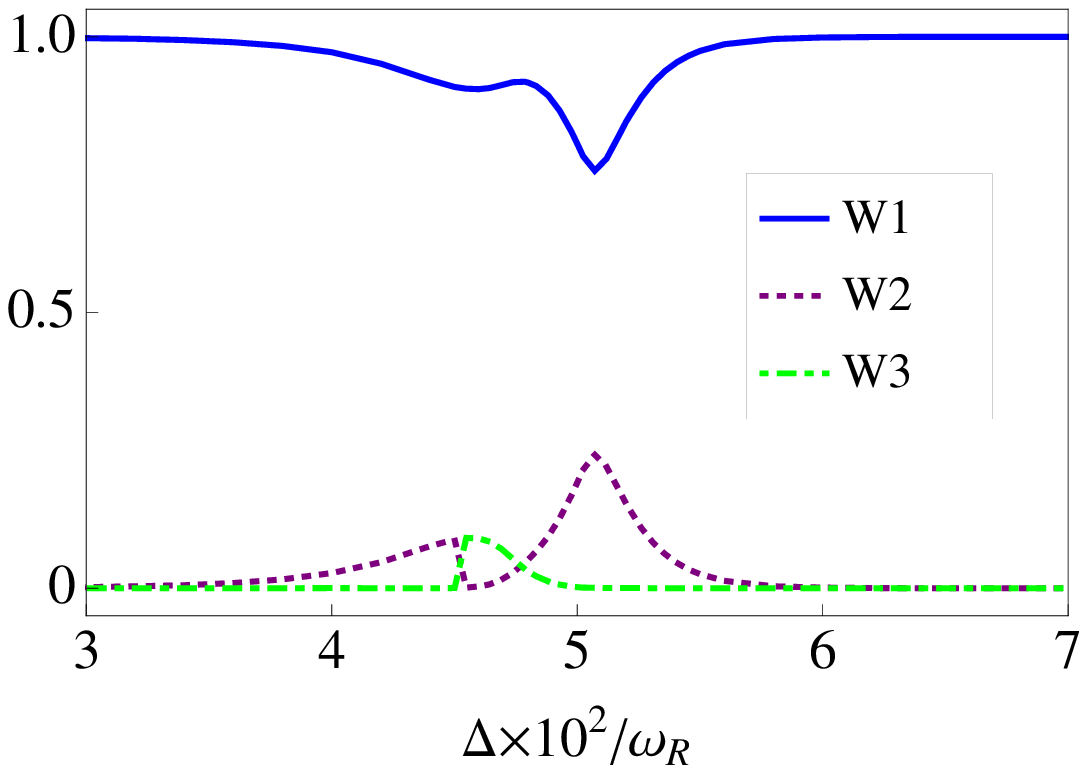}
\caption{Real parts of the eigenvalues $\lambda_i$ (panel (a)) and their weights $W_i$
(panel (b)) as function of the detuning. The results are obtained from the FPE method
with $\kappa = 7.2\times10^{-5}~\omega_{\rm R}$ and $g = 0.003~\omega_{\rm R}$.
}\label{fig:eigenk}
\end{figure}

For strong coupling or weak damping of the resonator the linewidth is enhanced
around the transition regime.
In this case, more than one eigenvalue contribute to the spectrum, and it has no longer 
a Lorentzian form.
Since the eigenvalues with small real parts are of importance for the linewidth,
we consider the first three, $\{\lambda_k\}$ (ordered with growing real part). 
Their weights defined as
\begin{eqnarray}
 W_i = \frac{|\alpha_i|^2}{\sum_k|\alpha_k|^2},
\end{eqnarray}
determine how much the corresponding eigenfunctions contribute to the spectrum.
As shown in Figs. \ref{fig:eigeng} and \ref{fig:eigenk}, when approaching  the transition regime,
the weight of the first eigenvalue, which dominates in the non-lasing and deep lasing regimes,
decreases, while the second or even the third eigenvalues becomes important.

\end{document}